\documentclass[pra,twocolumn,showpacs]{revtex4}
\pdfoutput=1
\usepackage{amsmath}
\usepackage{amssymb}
\usepackage{graphicx}

\newcommand{\ket}[1]{\mbox{$\vert #1 \rangle$}}

\begin{document}
\title{Imprinting Patterns of Neutral Atoms in an Optical Lattice using Magnetic Resonance Techniques}
    \author{Micha\l\ Karski}\email{karski@uni-bonn.de}
    \author{Leonid F\"orster}
    \author{Jai-Min Choi}
    \author{Andreas Steffen}
    \author{Noomen Belmechri}
    \author{Wolfgang Alt}
    \author{Dieter Meschede}
    \author{Artur Widera} \email{widera@uni-bonn.de}
    \address{Institut f\"ur Angewandte Physik, Universit\"at Bonn,
    Wegelerstr.~8, D-53115 Bonn, Germany}
\date{\today}

\begin{abstract}
We prepare arbitrary patterns of neutral atoms in a one-dimensional (1D) optical lattice with single-site precision using microwave radiation in a magnetic field gradient.
We give a detailed account of the current limitations and propose methods to overcome them.
Our results have direct relevance for addressing of planes, strings or single atoms in higher dimensional optical lattices for quantum information processing or quantum simulations with standard methods in current experiments. Furthermore, our findings pave the way for arbitrary single qubit control with single site resolution.
\end{abstract}

\maketitle

\section{Introduction}
Neutral atoms trapped in optical lattices form a promising paradigm of quantum simulation \cite{bloch_many-body_2008} and quantum information processing \cite{bloch_quantum_2008}. Numerous proposals in these fields require the ability to coherently address and manipulate atoms on a single site of the lattice. This ability, however, still poses a challenge in these systems. While detection of atoms in optical lattices with single site resolution has been reported \cite{scheunemann_resolving_2000, nelson_imaging_2007, gericke_high-resolution_2008, karski_nearest-neighbor_2009, bakr_quantum_2009}, single site manipulation in lattices with a site separation in the optical wavelength domain has so far only been demonstrated by removing atoms using a focussed electron beam \cite{wurtz_experimental_2009}.

A convenient method for spatially resolved coherent manipulation and detection of atoms using only global techniques originates from nuclear magnetic resonance (NMR). Originally developed for solid state systems, the basic tools and concepts have found their way also into quantum optics experiments using cooled trapped atoms \cite{vandersypen_nmr_2005}. Achievements include control of internal states on a micrometer scale \cite{schrader_neutral_2004} or robust global control of atomic samples \cite{rakreungdet_accurate_2009}. In large systems, where a Bose-Einstein condensate is loaded into an optical lattice, periodic patterned loading was achieved by using superlattice potentials \cite{peil_patterned_2003, folling_direct_2007}. In an array of double wells, recently one well out of each double well system could be selectively manipulated by an optically induced effective magnetic field, while the qubit was stored in field insensitive internal states \cite{lundblad_field-sensitive_2009}.

Here, we demonstrate the use of NMR techniques to prepare arbitrary patterns of atoms in a 1D optical lattice with single-site precision, which can be used as a starting point for quantum information processing, and can be extended to higher dimensional systems. These techniques overcome several restrictions posed on atom string preparation using moving optical lattices \cite{miroshnychenko_precision_2006}, such as the limited resolution on the order of the beam diameter or the limited selectivity for closely spaced atoms. In principle, the NMR techniques can be realized with standard methods already used in current quantum gas experiments and can be extended to yield arbitrary quantum state manipulation of qubits with single site resolution. We discuss the challenges of this extension in the context of our 1D optical lattice.

\section{Experimental Setup}
\subsection{Cooling and Trapping of Atoms}
We capture neutral Caesium (Cs) atoms from the background gas in a three-beam magneto-optical trap (MOT).
The atoms are transferred into a far detuned standing wave optical dipole trap (1D optical lattice) formed by the interference of two counterpropagating laser beams with a wavelength of $\lambda_\mathrm{lat} = 866\,$nm. By superposing the two traps for a duration of 500\,ms and subsequently switching off the MOT, depending on the parameters, between a single and up to 100 atoms can be loaded into the optical lattice. For our parameters, atoms occupy a region of approximately $100\,\mu$m length of the lattice (see Fig.\ref{fig:GradientSpectroscopy}), corresponding to roughly 200 lattice sites.

During the loading and fluorescence imaging phases (see below) the dipole trap has a depth of $k_B \times 0.4\,$mK, whereas for manipulation of the atoms with microwaves in a magnetic field gradient it is adiabatically lowered within 50\,ms to a depth of $k_B \times 80\,\mu$K. Here, in each dipole trap beam, approximately $20\,$mW of laser beam power are focussed down to a waist radius of $20\,\mu$m. This corresponds to vibrational frequencies of $\omega_\mathrm{ax} = 2\pi \times 115\,$kHz along the optical lattice axis (axial direction, $z$-axis) and $\omega_\mathrm{rad} = 2\pi\times 1.2\,$kHz perpendicular to the optical lattice axis (radial direction, $x y-$plane). The atoms have a final temperature of $k_B\times10\,\mu$K, corresponding to mean vibrational quantum numbers of $\bar{n}_\mathrm{ax} = 1.2$ axial and $\bar{n}_\mathrm{rad} = 200$ .

The atomic sample is probed by fluorescence imaging. We typically illuminate the trapped atoms between 200\,ms and 1\,s by a near-resonant optical molasses using the MOT laser beams, and image the fluorescence onto an EMCCD (electron multiplying CCD) camera. The particular exposure time is chosen depending on the required precision for the determination of atomic positions \cite{karski_nearest-neighbor_2009}. For our imaging system and the fluorescence wavelength of 852\,nm the diffraction limit is $1.8\,\mu$m $\approx 4\times \lambda_\mathrm{lat}/2$. However, for a sparsely filled lattice, even in unresolved clusters of less than eight atoms, numerical post-analysis allows high-precision, real-time determination of the atomic position with down to nearest neighbor distances \cite{karski_nearest-neighbor_2009}.

\begin{figure}
\centering
  \includegraphics[scale=0.45]{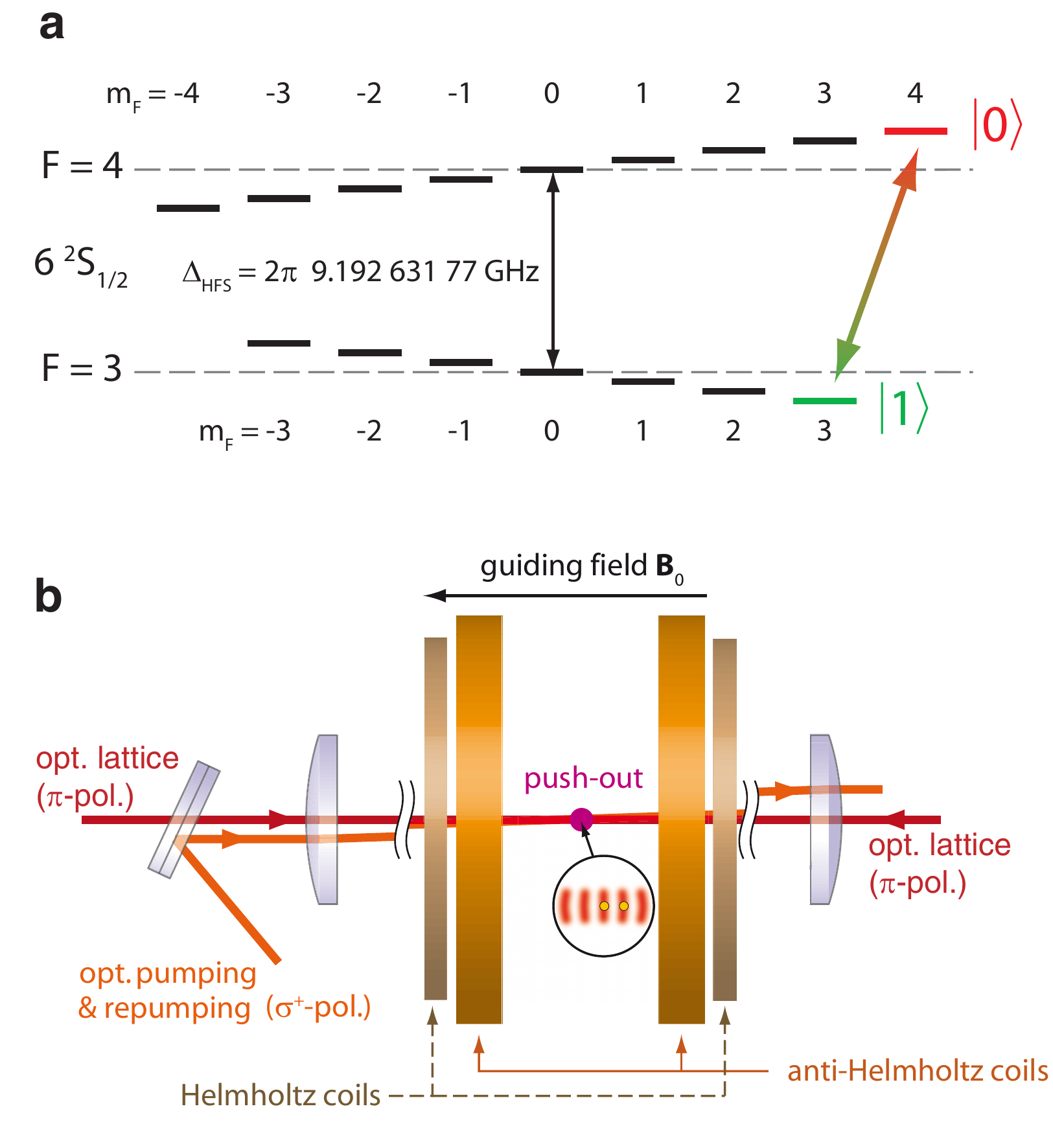}\\
  \caption{({\bf a}) Zeeman splitting of magnetic sublevels in the $6_{}^{2}S_{1/2}^{}$ ground state manifold of caesium. The degeneracy of the magnetic sublevels is lifted by an external guiding field due to the linear Zeeman effect. State ${|0\rangle = |F=4,m_{\mathrm{F}}^{}=4\rangle}$ and state ${|1\rangle = |F=3,m_{\mathrm{F}}^{}=3\rangle}$ define the states of the qubit. ({\bf b}) Geometrical arrangement of laser beams and coils producing the magnetic fields needed for preparation and detection of arbitrary patterns of atoms.}\label{fig:OpticalPumping}
\end{figure}
\subsection{Internal State Manipulation and Detection}
We initialize the atoms in state $\ket{0} \equiv \ket{F=4, m_F=+4}$ by optical pumping with a $\sigma_{}^{+}$-polarized laser beam, which is resonant with the ${F=4\rightarrow F'=4}$ transition. Here $F$ is the total atomic angular momentum and $m_F$ its projection onto the quantization axis, determined by a weak ($\vert \mathbf{B}_0 \vert = 3\,$G) guiding magnetic field along the $z$ axis. An equally polarized repumping laser beam, frequency stabilized on the ${F=3\rightarrow F'=4}$ transition, is used to transfer the atoms back to the optical pumping cycle, whenever they decay to the ${F=3}$ ground state.

As information is stored in our experiment in the hyperfine states \ket{0} and $\ket{1} \equiv \ket{F=3, m_F=+3}$, we have to selectively probe the populations of these states. This is done using the so-called ``push-out'' technique \cite{kuhr_coherence_2003}, which removes the atoms in ${F=4}$ (including state $|0\rangle$) from the optical lattice while leaving those in ${F=3}$ unaffected. For this, we apply an intense laser beam operating on the ${F=4\rightarrow F^\prime=5}$ transition, perpendicular to the optical lattice axis. The power ${P_{\mathrm{push}}^{}=40\,\mu\mathrm{W}}$ and the pulse duration ${\tau_{\mathrm{push}}^{}=250\,\mu\mathrm{s}}$ of the push-out beam are optimized so that its radiation pressure force overcomes the radial dipole force and quickly pushes the atoms in ${F=4}$ out of the lattice within less than half a radial oscillation period. Thereby, we largely prevent off-resonant excitations to ${F'=4}$, from where the atoms can spontaneously decay into the ${F=3}$ ground state, which would cause erroneous detection of $F=3$.
We verified that the mean survival probability of atoms prepared in ${F=4}$ is smaller than $1\%$, whereas for atoms prepared in ${F=3}$, it is larger than $99\%$, imposing lower limits of the push-out efficiency.

The two internal states \ket{0} and \ket{1} are coupled using microwave radiation around 9.2\,GHz. The microwave field is created by mixing a fixed frequency signal from a phase locked dielectric resonator oscillator (PLDRO) around 9.0\,GHz with a tunable frequency from a signal generator around 200\,MHz in an upconverter. Both sources are locked onto a Rubidium clock. The resulting sum frequency is amplified up to 12\,W and directed onto the atoms using a commercial waveguide that supports linear polarization, while the carrier and difference frequencies are suppressed by at least -30\,dB.

We thereby achieve Rabi frequencies of approximately $\Omega_0 = 2 \pi \times 60\,$kHz for the $\ket{0} \leftrightarrow \ket{1}$ transition. The coherence time of the system has been measured by a Ramsey-type pulse sequence to be $T_2 \approx 200\,\mu$s and can be increased to $T_2^\ast \approx 0.8\,$ms by a spin echo sequence.

\subsection{Magnetic Gradient Field}

In order to create a spatially varying resonance frequency along the optical lattice, we apply a quadrupole magnetic field $\mathbf{B}_\mathrm{quad}(\mathbf{r}, I)$. The field originates from two coils in anti-Helmholtz configuration, where the symmetry axis of the coils coincides with the optical lattice axis. At the center of the trapping region, the axial field gradient along the $z$-axis is thus twice as large as the gradient in the $xy-$plane. The magnetic field in the trapping region including the offset field $B_0\, \hat{\mathbf{e}}_z$ then reads
\begin{equation}\label{eq:Bfield}
    \mathbf{B}(\mathbf{r},I) = \left ( \begin{array}{c} 0 \\ 0 \\B_0 \end{array}  \right ) +
        B^\prime(I)\, \left( \begin{array}{c} -x/2 \\ -y/2 \\ z \end{array} \right),
\end{equation}
where $B^\prime(I)$ denotes the magnitude of the gradient field along the $z$-axis for a chosen current $I$ running through the coils. Due to the linear Zeeman effect, the  ${|0\rangle\leftrightarrow|1\rangle}$ transition frequency depends linearly on the modulus of $\mathbf{B}(\mathbf{r},I)$, yielding

\begin{eqnarray}
\omega_{0}(\mathbf{r},I) &=\Delta_{\mathrm{HFS}}+\gamma|\mathbf{B}(\mathbf{r},I)|\nonumber\\
&\approx\Delta_{\mathrm{HFS}}+\gamma\left(B_{0}+B'(I)z + \frac{B'(I)^2}{8B_{0}}\rho^{2}\right)\nonumber\\
&=\Delta_{\mathrm{HFS}}+\delta_{0}+\omega'(I) z+\frac{\omega'(I)^{2}}{8\delta_{0}}\rho^{2}\,,\label{eq:transition-frequency-gradient}
\end{eqnarray}
where the gyromagentic ratio ${\gamma=(3g_{3}-4g_{4})\mu_{B}/\hbar} \approx 2 \pi \times 2.5\,$MHz/G, ${\delta_{0}=\gamma B_{0}}$ denotes the contribution of the guiding field, ${\omega'(I)=\gamma B'(I)}$ the position-dependent shift and ${\rho=\sqrt{x^2+y^2}}$ the radial distance from the axis of symmetry of the coils. Finally, $|\mathbf{B}(\mathbf{r},I)|$ has been approximated to second order in $\rho$, valid for ${(B_{0}+B'(I)z)^{2}\gg B'(I)^{2}\rho^{2}/4}$. Note that, due to misalignment, the optical lattice axis can be radially offset by a small amount $\rho_{0}$ relative to the axis of symmetry. While the linear dependency of $\omega_{0}(z,\rho,I)$ on the axial position $z$ is still maintained, this offset imposes a significant obstacle to single site addressability, as is shown in Section \ref{sec:RadialOffset}.

From a calibration measurement tracking the shift of the microwave frequency in the lattice when certain magnetic gradient fields have been applied (see next section), we infer the position dependent frequency shift of

\begin{equation}\label{eq:gradient-frequency-current-calibration}
\frac{\omega'(I)}{2\pi I} = (291\pm 2)\frac{\mathrm{Hz}}{(\lambda/2)\,\mathrm{A}} = (671\pm 3)\frac{\mathrm{Hz}}{\mu\mathrm{m}\,\mathrm{A}}\,.
\end{equation}

This frequency shift corresponds to a strength of magnetic field gradient along the lattice axis of ${B'(I)/I=-(274\pm 1)\,\mu\mathrm{G}/(\mu\mathrm{m}\,\mathrm{A})}$. For the maximum current of 45\,A, we find a frequency separation for two neighboring lattice sites of $\omega^\prime/ (2 \pi) \,\times \lambda/2=  13\,$kHz, corresponding to a magnetic field gradient of approximately 120\,G/cm.

\section{Single Particle Operations with Spatially Varying Resonance Condition}
\subsection{Position Dependent Spectroscopy}
\begin{figure}
    \centering
  \includegraphics[scale=0.5]{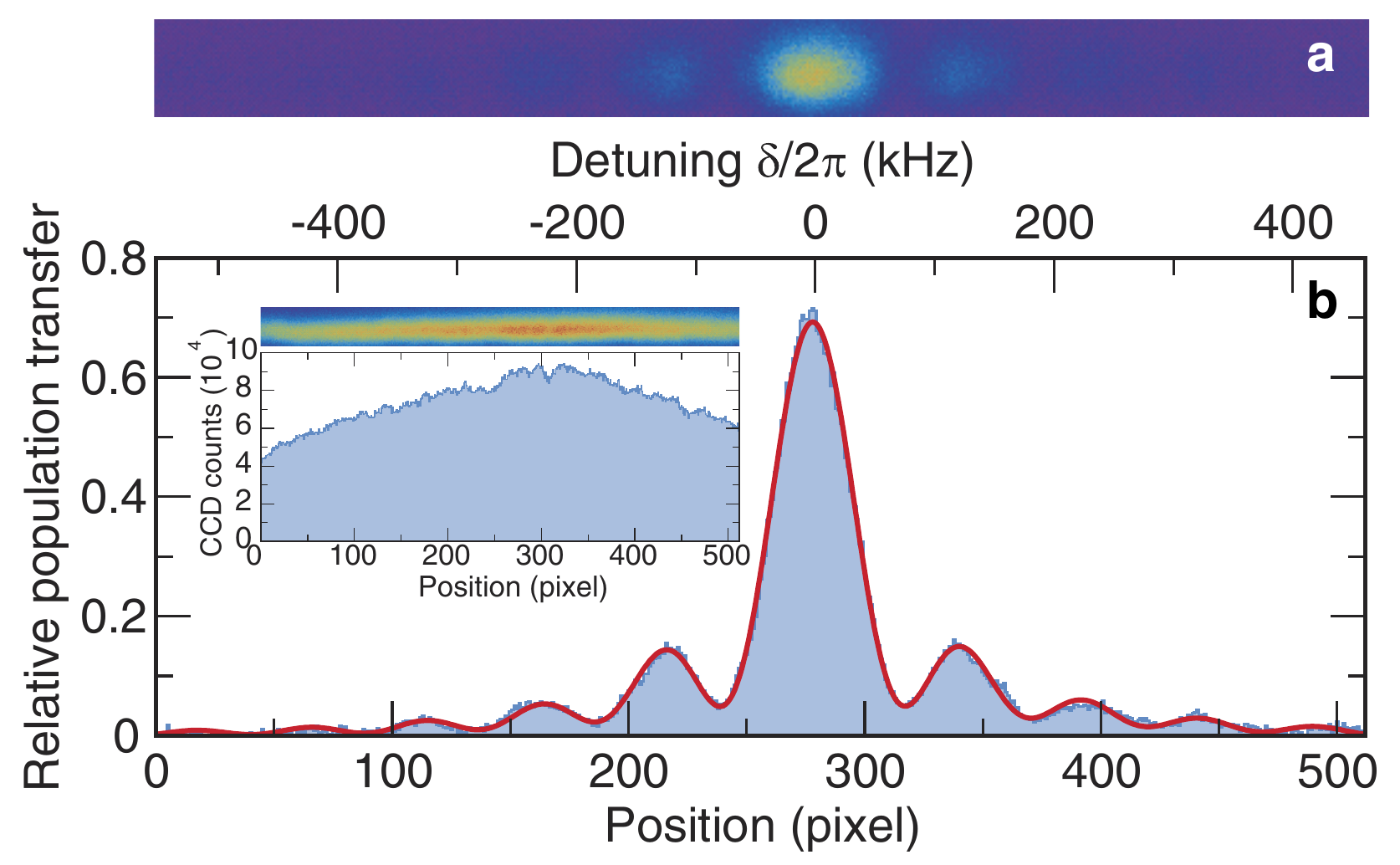}\\
  \caption{(a) Superposition of 50 fluorescence images after a rectangular microwave pulse has been applied to a filled optical lattice in a magnetic gradient field to flip the internal state from \ket{0} to \ket{1} and a subsequent laser pulse has removed atoms in \ket{1}. (b) Vertically binned intensity distribution of the images shown in (a). The solid line shows a fit with the expected sinc-like frequency profile. The insert shows the atom distribution without application of the microwave pulse and push-out laser.}\label{fig:GradientSpectroscopy}
\end{figure}

When a magnetic field gradient is applied during the microwave pulse, the atomic resonance condition is fulfilled only in small regions along the lattice axis. In order to illustrate this, we apply a rather weak magnetic field gradient of $B^\prime = 27.4\,$G/cm. For an atomic sample which has been initialized to the \ket{0} state, a subsequent microwave pulse with rectangular amplitude envelope of duration $t_\mathrm{pulse} = 10\,\mu$s and a center frequency of $9.2$\,GHz is applied. As a consequence, the frequency spectrum with its characteristic sinc-type side-peaks is mapped onto the position distribution of atoms in \ket{1}. This is directly observable in the fluorescence images, taken after the push-out has been applied, see Fig.~\ref{fig:GradientSpectroscopy}. The averaged intensity distribution of several images thus directly illustrates the shape of the microwave pulse in frequency domain.
Microwave spectroscopy in position space is therefore a useful tool, which immediately and quickly reveals information encoded in the frequency domain with a high resolution, controlled by the strength of the field gradient. This method is much faster than a usual scan in frequency across the resonance in a homogeneous field to map out the full spectrum. Further, using microwave pulses with a narrow-band spectrum, it allows us to monitor the evolution of the transition frequency arising from changes and drifts of experimental parameters from shot to shot. We stress that the shape of the microwave spectrum in position space is broadened due to the optical imaging when features in frequency space become smaller than the optical diffraction limit. A further broadening may exist due to a radial offset of the lattice axis with respect to the axis of symmetry of the coils. This effect, however, becomes only significant for narrow-band pulses with a high spectral selectivity as we discuss in Section \ref{sec:RadialOffset}.

\subsection{Patterned Atomic String Preparation}
The preparation of convenient initial atom configurations is an important first step for applications of neutral atoms for quantum information technology, such as the creation of entangled states through coherent collisions; the tailoring of atom strings for efficient interaction with the field of a high-finesse resonator; or the extraction of a selected plane or string of atoms from a Mott-insulating state of atoms \cite{greiner_quantum_2002, stferle_transitionstrongly_2004, spielman_mott-insulator_2007}. The concepts presented above provide a toolbox which offers this capability: In presence of a magnetic field gradient, only those lattice sites remain occupied after the state-selective push-out, at which atoms have been shifted resonant to the microwave $\pi$-pulse.

In order to generate a predefined pattern structure a pulse train of $N$ successively applied $\pi$-pulses with different frequencies $\omega_{i}$ is used instead of a single pulse. By incorporating the periodicity of the optical lattice using

\begin{equation}
	\frac{|\omega_{i}-\omega_{j}|}{\omega'(I)\lambda/2}\in\mathbb{N}\quad\mathrm{with}\quad i,j=1,\ldots,N\,,\,i\not=j\,,
\end{equation}
these frequencies define the pattern structure. To ensure a high selectivity in the position-dependant population transfer, we use the gradient providing the maximum frequency shift of  $\omega^\prime (I) /(2\pi) \times (\lambda/2) = 13\,\mathrm{kHz}$. Furthermore, Gaussian $\pi$-pulses with a $1/\sqrt{e}$ spectral half-width down to ${\sigma_{\omega}/2\pi=6\,\mathrm{kHz}}$ are employed rather than rectangular pulses in order to suppress side lobes. For this maximum available frequency shift, these pulses are in principle capable of manipulating individual atoms with almost single site resolution. To increase the selectivity in the preparation of the pattern structures, we optionally use several iterations of state initialization, application of $\pi$-pulses and the state-selective push-out.
\begin{figure}
    \centering
  \includegraphics[scale=1]{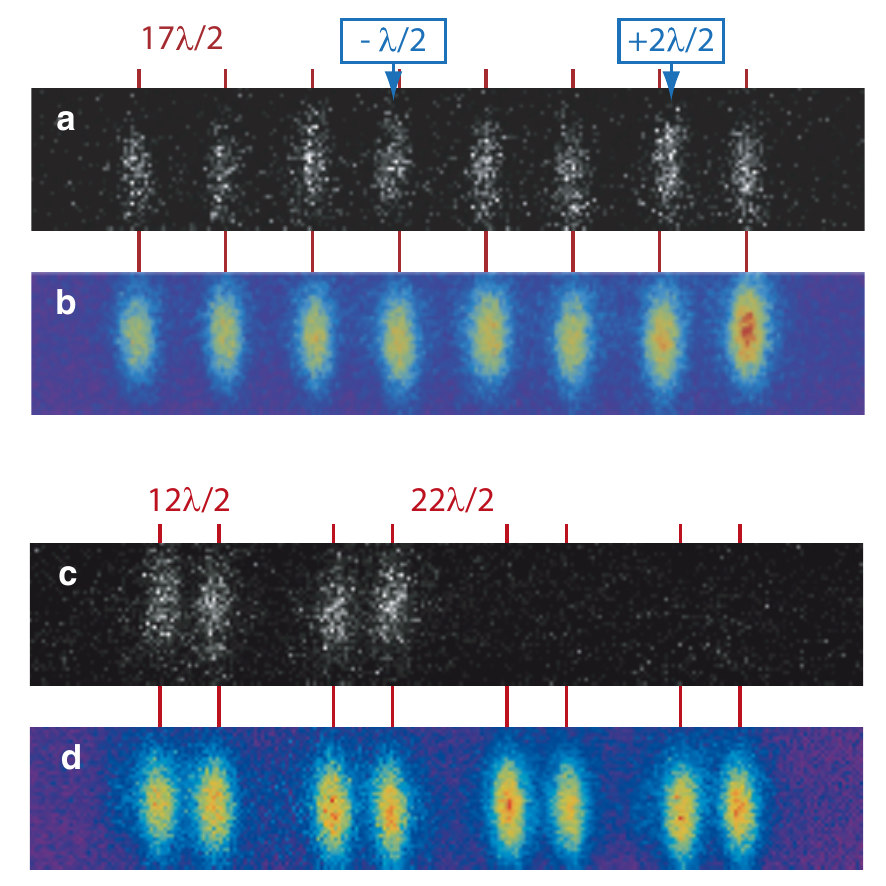}\\
  \caption{Strings of atoms in an optical lattice: Greyscale images show a single image acquired after application of the patterning sequence for ({\bf a}) a predefined string of equidistant atoms with separations of $17\lambda/2$, and ({\bf c}) for a predefined string of four "atom pairs" with an atom separation of 12 lattice sites within each pair. Corresponding averages over 50 acquired images are shown as false color images, ({\bf b}) and ({\bf d}), respectively. Atoms deviating from predefined positions are indicated by an arrow, the deviation by a boxed value. Missing atoms in ({\bf b}) are mostly attributed to the limited preparation efficiency (see text).}\label{fig:atom-pattern}
\end{figure}

In Figure \ref{fig:atom-pattern}, two exemplarily generated patterns are shown: A string of eight equidistant atoms (Fig.~\ref{fig:atom-pattern}(a,b)), for instance, is ideally suited for a quantum register. Such a register has previously been implemented using atoms randomly distributed over the lattice sites \cite{schrader_neutral_2004}, in which case the position of each atom needed to be determined and fed back to the microwave source prior to the quantum state manipulation of the atoms. For a predefined string of atoms, this feedback is not required, since the frequency of each individual atom of the string is defined by the preparation sequence itself.

Compared to rearranging atoms in crossed movable 1D lattices \cite{miroshnychenko_precision_2006,miroshnychenko_quantum_2006}, our method allows us to prepare strings of atoms separated by an arbitrary number of lattice sites, even down to two (see Fig.~\ref{fig:histogram-pairs-string}(a,b) below). These small separations are essential for an efficient implementation of controlled collisions of two individual atoms \cite{jaksch_entanglement_1999, brennen_quantum_1999, srensen_spin-spin_1999, mandel_controlled_2003} using, e.g., state-selective transport \cite{mandel_coherent_2003, karski_quantum_2009}.

\subsection{Limitations}
\subsubsection{Filling factor}
The first restriction of our method is given by the initial loading from a MOT. For high densities as in an optical lattice, light-induced collisions \cite{schlosser_sub-poissonian_2001, weber_analysis_2006} lead to high loss rates. In our case, any doubly occupied lattice site illuminated by the near-resonant molasses is quickly depleted \cite{forster_number-triggered_2006}. The resulting steady-state distribution of atoms in the lattice will show only either one or no atom, resulting in a filling factor of about $0.5$,  yielding a probability to find an atom at one selected lattice site of ${p_{\mathrm{a}} \approx 50\%}$. This probability imposes an upper limit for the efficiency of generating the entire pattern structure: For a pattern of $N$ atoms, the probability that all desired lattice sites are initially populated is given by ${p_{\mathrm{ini}}=p_{\mathrm{a}}^{N}}$, yielding ${p_{\mathrm{ini}}\approx 0.4\%}$ for a string of $8$ atoms. Since our atom detection is restricted to sparsely filled lattices, a detailed investigation of the loading process with high atom densities could not be performed so far. Further investigations aiming on the increase of the filling factor are required in the future.

\subsubsection{Selectivity of microwave pulses}
For an atom at a predefined lattice site, the selective microwave operation, a $\pi$-pulse for instance, should fulfill two conditions. First, it should operate on the selected atom with high fidelity. The selectivity of this $\pi$-pulse, in turn, specified by its spectral width and its shape in the frequency domain, determines to what extent neighboring atoms are affected by this pulse and whether neighboring atoms are completely removed from the lattice or remain trapped with a certain probability. For efficient preparation of pattern structures, this probability should by ideally zero.

However, the selectivity cannot be infinitely improved by increasing the pulse duration. We observe experimentally that for pulse durations longer than $15\,\mu$s the maximally achievable population transfer in the pulse center decreases together with the spectral width. We attribute this to decoherence stemming from, e.\,g.\ technical noise or inhomogeneous broadening due to radial motional dynamics in combination with the differential AC Stark shift due to the lattice light \cite{kuhr_analysis_2005}.
For this reason, we employ a simple technique, which at least for the generation of patterned structures presented in this work, effectively improves the selectivity of the Gaussian $\pi$-pulses. The underlying idea is to utilize the probabilistic and destructive nature of our state-selective push-out.

Suppose that a single application of the microwave spectroscopy sequence with a Gaussian $\pi$-pulse provides a spectrum given by
\begin{equation}\label{eq:gauss-pulse-spectrum}
	P_{|1\rangle}(\delta) =P_{\mathrm{max}}\exp\left(-\frac{\delta^2}{2\sigma_{\omega}^{2}}\right)\,,
\end{equation}

\begin{figure}[t!]
	\centering
	\includegraphics[width=1\columnwidth]{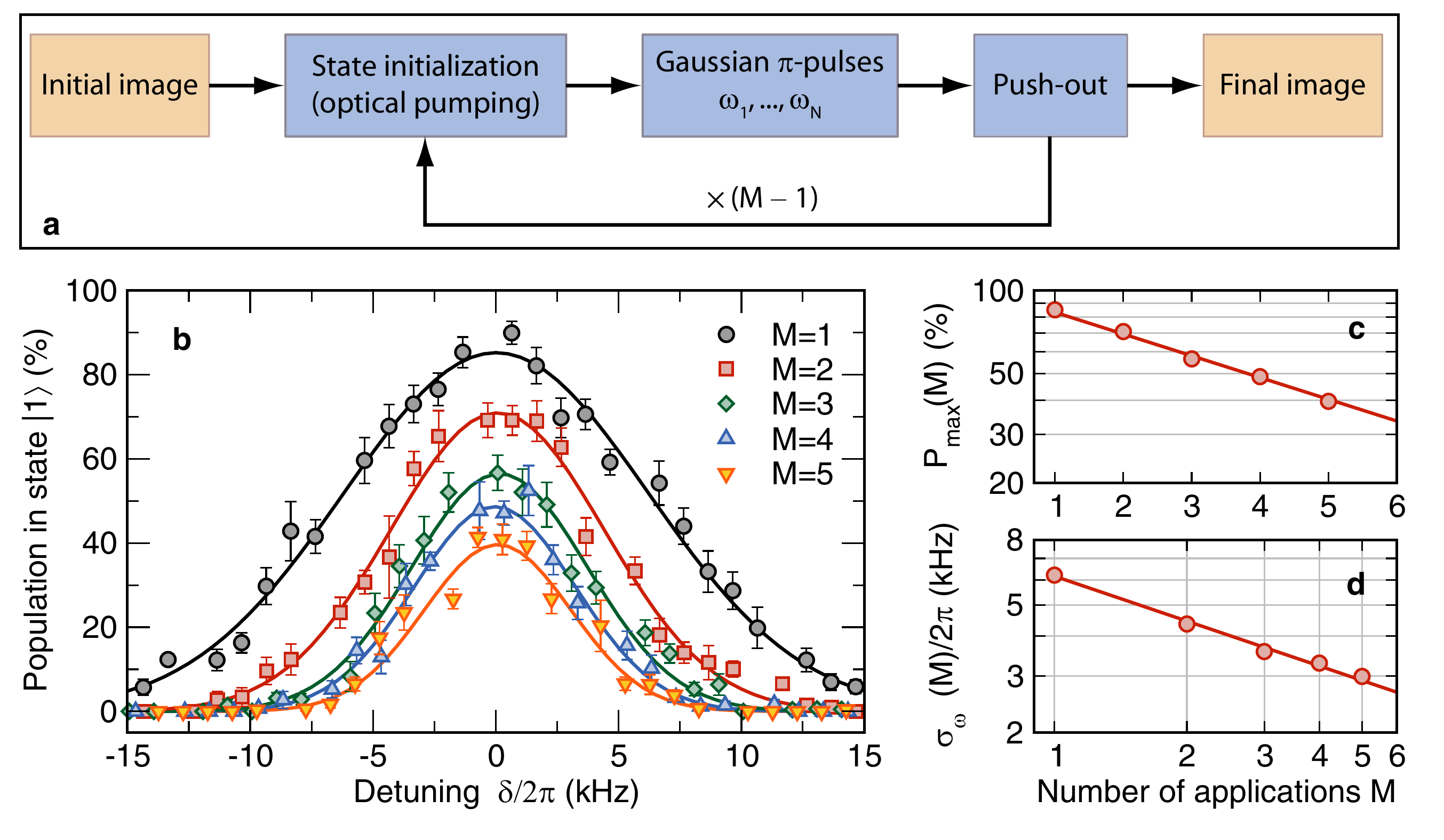}
	\caption{({\bf a}) Sequence for preparation of pattern structures involving multiple application of the inner loop of state transfer and push-out (blue shaded blocks). ({\bf b}) Microwave spectra in frequency domain for different numbers of applications $M$ of the inner loop. Solid lines show Gaussian fits, from which ({\bf c}) the dependency of the inferred maximum transfer efficiency $P_{\mathrm{max}}(M)$ (logarithmic plot) and ({\bf d}) the $1/\sqrt{e}$ spectral half-width $\sigma_{\omega}(M)$ (double-logarithmic plot) on the number of applications is inferred. Both dependencies perfectly agree with the expected trend.} \label{fig:multi-push-out}

\end{figure}

with a certain maximum population transfer $P_\mathrm{max}$. Then, by repeatedly applying the inner loop comprising state initialization, application of the $\pi$-pulse and the push-out by a total of $M$ times (see Fig.~\ref{fig:multi-push-out}), we expect again a Gaussian shaped spectrum

\begin{eqnarray}
	P_{|1\rangle}^{M}(\delta) &=&\left[P_{\mathrm{max}}\exp\left(-\frac{\delta^2}{2\sigma_{\omega}^{2}}\right)\right]^{M} \\ \nonumber &=&(P_{\mathrm{max}})^{M}\exp\left(-\frac{\delta^2}{2{(\sigma_{\omega}/\sqrt{M})}_{}^{2}}\right)\,,
\end{eqnarray}
with a $1/\sqrt{e}$ spectral half-width ${\sigma_{\omega}(M)=\sigma_{\omega}/\sqrt{M}}$, however, with a reduced maximum population transfer ${P_{\mathrm{max}}(M)=(P_{\mathrm{max}})^{M}}$ .

In Figure \ref{fig:multi-push-out}, the scaling behavior of both quantities is exemplarily shown for a Gaussian pulse with ${\sigma_{t}=20\,\mu\mathrm{s}}$, where the spectra have been taken in the frequency domain rather than by position dependent spectroscopy. The scaling is perfectly reproduced by the measured data, for both quantities inferred from a fit. Comparing the performance of a sequence repeating the inner loop with a less selective $\pi$-pulse of ${\sigma_{t}=15\,\mu\mathrm{s}}$ by a total of two or three times with a single application of a more selective pulse of ${\sigma_{t}=20\,\mu\mathrm{s}}$, we conclude that, for the efficient generation of patterns, the repeated application is more advantageous.

\begin{figure}[t!]
	\centering
	\includegraphics[scale=0.75]{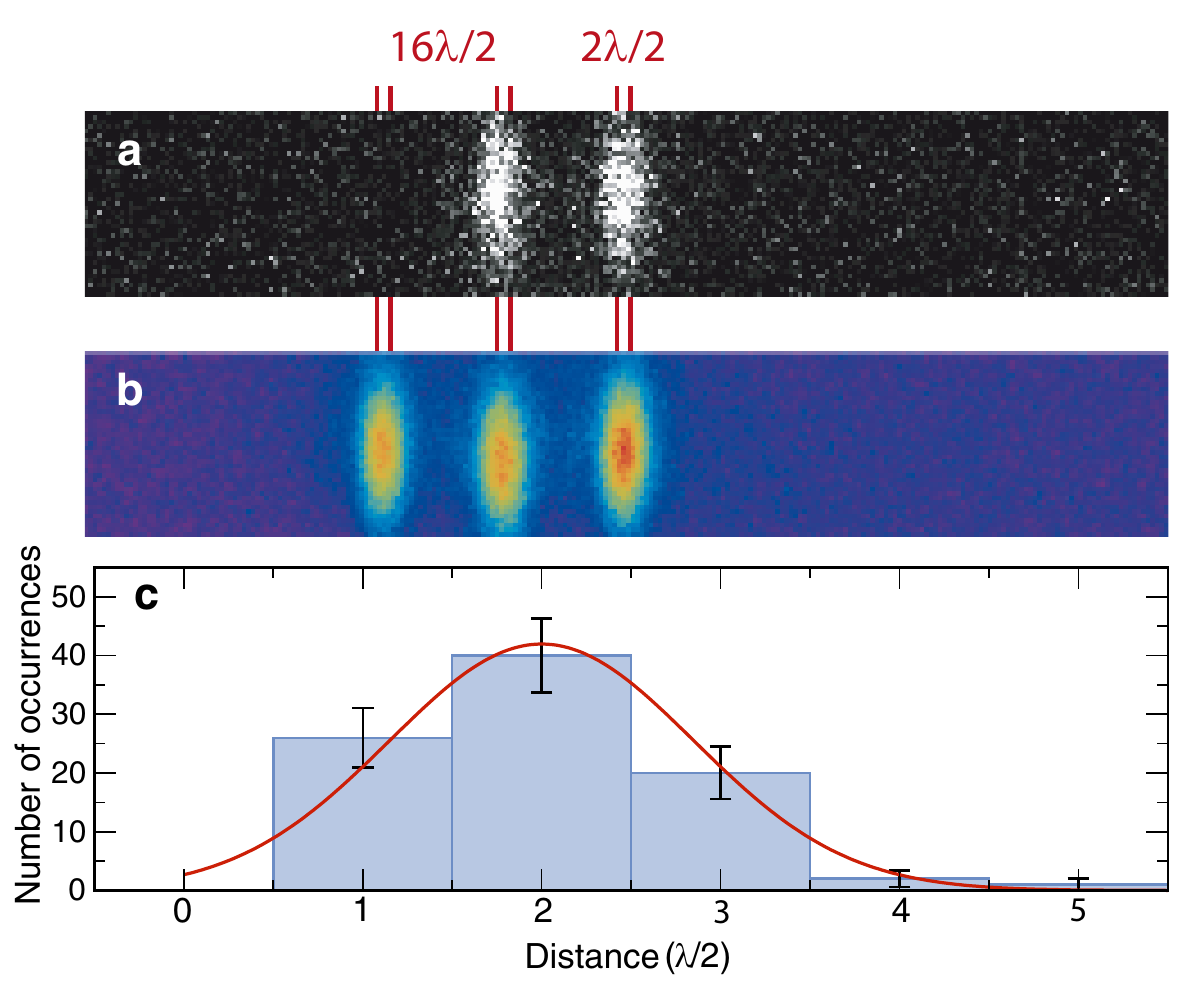}
	\caption{Pattern of three trapped atom pairs with nearest neighbor distance ({\bf a}) in a single and ({\bf b}) the average of several fluorescence images, similar to the patterns shown in Fig.~\ref{fig:atom-pattern}. ({\bf c}) Histogram of distances between two atoms forming the pairs. The solid line shows a Gaussian fit to the histogram, from which the $1/\sqrt{e}$ half-width of the distances and the selectivity region of a single atom are inferred.} \label{fig:histogram-pairs-string}

\end{figure}

To investigate the efficiency, the resolution and possible imperfections of our patterning method, we generate a pattern of three pairs of next nearest neighbors, see Fig.~\ref{fig:histogram-pairs-string}. Each atom pair is separated by $16$ lattice sites from another, to reliably determine the distances of the atoms using numerical post-processing as presented in Ref.~\cite{karski_nearest-neighbor_2009}. The internal sequence core employs a pulse train of six subsequent applied Gaussian $\pi$-pulses with ${\sigma_{t}=20\,\mu\mathrm{s}}$.
The inner loop is repeated by a total of ${M=2}$ times. Consequently, we expect a maximum population transfer of ${P_{\mathrm{max}}(2)=(71\pm 2)\%}$ at the predefined sites. The corresponding $1/\sqrt{e}$ half-width in position space is $\sigma_{z,\mathrm{exp}}(2)=(0.34\pm 0.01)\lambda/2\,$,
specifying the selectivity region for a single atom.

The entire patterning sequence is repeatedly recorded $500$ times. From the final fluorescence images of the atoms, each acquired with an exposure time of $800\,\mathrm{ms}$, the positions and the distances of the atoms are determined.
For our parameters, the number of pairs correctly prepared with the predefined distance is expected to be $N_{\mathrm{exp}} \approx 170\pm17$.
In Figure \ref{fig:histogram-pairs-string} a histogram of the measured distances between atoms forming the pairs is shown. It reveals a Gaussian shaped distribution centered at the predefined separation of two lattice sites. From the histogram, we infer ${N_{\mathrm{meas}}=40}$ correctly prepared pairs. From the fitted width of the distribution $\sigma_{\mathrm{dist}}$ we infer a selectivity region of a single atom as $\sigma_{\mathrm{meas}}=\sigma_{\mathrm{dist}}/\sqrt{2}=(0.60\pm 0.06)\,\lambda/2\,$.

Both, the number of correctly prepared pairs and the $1/\sqrt{e}$ half-width of the selectivity region deviates from the expected values. These deviations can be attributed to the axial drift of the optical lattice and the radial offset of the lattice axis relative to the axis of symmetry of the coils, as discussed below.

\subsubsection{Axial drift of the lattice}
\begin{figure}
  \centering
  \includegraphics[width=1\columnwidth]{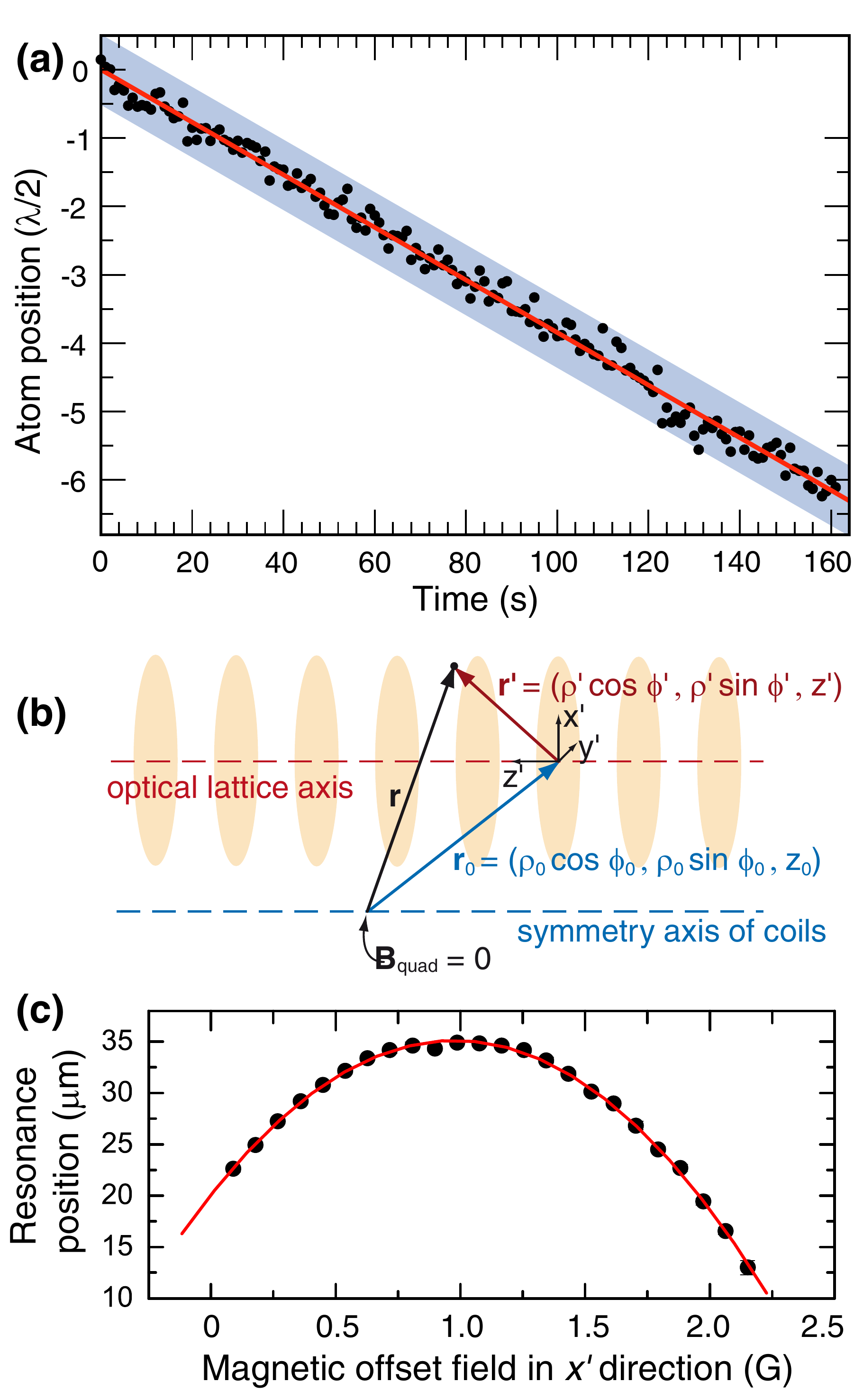}\\
  \caption{(a) Drift of the optical lattice with respect to the imaging system over long times. (b) New coordinate system for an offset between the symmetry axes of optical lattice and gradient field. (c) Measured quadratic dependence of the position of the microwave resonance frequency with increased offset $\mathbf{r}_0$ along the $z$-direction. The offset is changed by applying a homogeneous offset field along the $z$direction.}\label{fig:RadialAxialShift}
\end{figure}

The standing wave field of the optical lattice is susceptible to drifts and fluctuations of the optics, effectively changing the position of the potential minima over the course of a measurement.
To infer the drift of the optical lattice relative to the imaging optics, we track the position of a single atom over a long time period from successively acquired images with $1\,\mathrm{s}$ exposure time each, see Fig.~\ref{fig:RadialAxialShift}a.
We observe small fluctuations of the atom position around an approximately linear trend, indicating a slow drift of the optical lattice of about ${10\,\mathrm{nm}/\mathrm{s}}$. We attribute this drift to thermal expansion of the optical table and opto-mechanics of the lattice.

Because the total measurement time is much longer than the time interval in which the lattice axially drifts over a distance of $\lambda/2$, the effect of this  drift on the spectrum in position space can be described by a convolution equation

%
\begin{equation}
P_{|1\rangle,\mathrm{drift}}(z) =\int\limits_{-\frac{\lambda}{4}}^{+\frac{\lambda}{4}} \frac{2}{\lambda} \, P_{|1\rangle}(z-z')\mathrm{d}z'\,,
\end{equation}
where we have included the fact that only drifts modulo $\lambda/2$ are distinguishable and relevant. By solving the convolution equation, we infer a ``drift-free'' $1/\sqrt{e}$ half-width of the selectivity region of $\sigma_{z,\mathrm{df}}=(0.52\pm 0.07)\lambda/2$. This value explains part of the deviation between measurement and expectation mentioned above.

\subsubsection{Radial offset of the lattice}\label{sec:RadialOffset}
According to Eq.~(\ref{eq:transition-frequency-gradient}) the transition frequency changes quadratically with increasing radial offset between the coil axis and the lattice axis. In the following, we calculate the effect of a radial offset of the lattice axis with respect to the axis of symmetry of the coils on the maximum population transfer and the selectivity region of a Gaussian $\pi$-pulse as used above.

To calculate the position dependent detuning, the origin of the lattice coordinate system for $\mathbf{r}^\prime$, which so far was located at the center of the lattice site where the magnetic quadrupole field vanishes, is shifted by the offset of the gradient field $\mathbf{r}_0$ (see Fig.~\ref{fig:RadialAxialShift}b). For a position vector $\mathbf{r}$ in this new coordinate system, the spatial detuning is given by

\begin{equation}\label{eq:detuning-field-gradient}
\delta(\mathbf{r}^\prime, \mathbf{r}_{0})=\omega_{0}(\mathbf{r}^\prime + \mathbf{r}_{0},I) - \omega_{0}(\mathbf{r}_{0},I)\,,
\end{equation}
with $\omega_{0}(\mathbf{r},I)$ from Eq.~(\ref{eq:transition-frequency-gradient}).

\begin{figure*}[t!]
	\centering
	\includegraphics[scale=0.6]{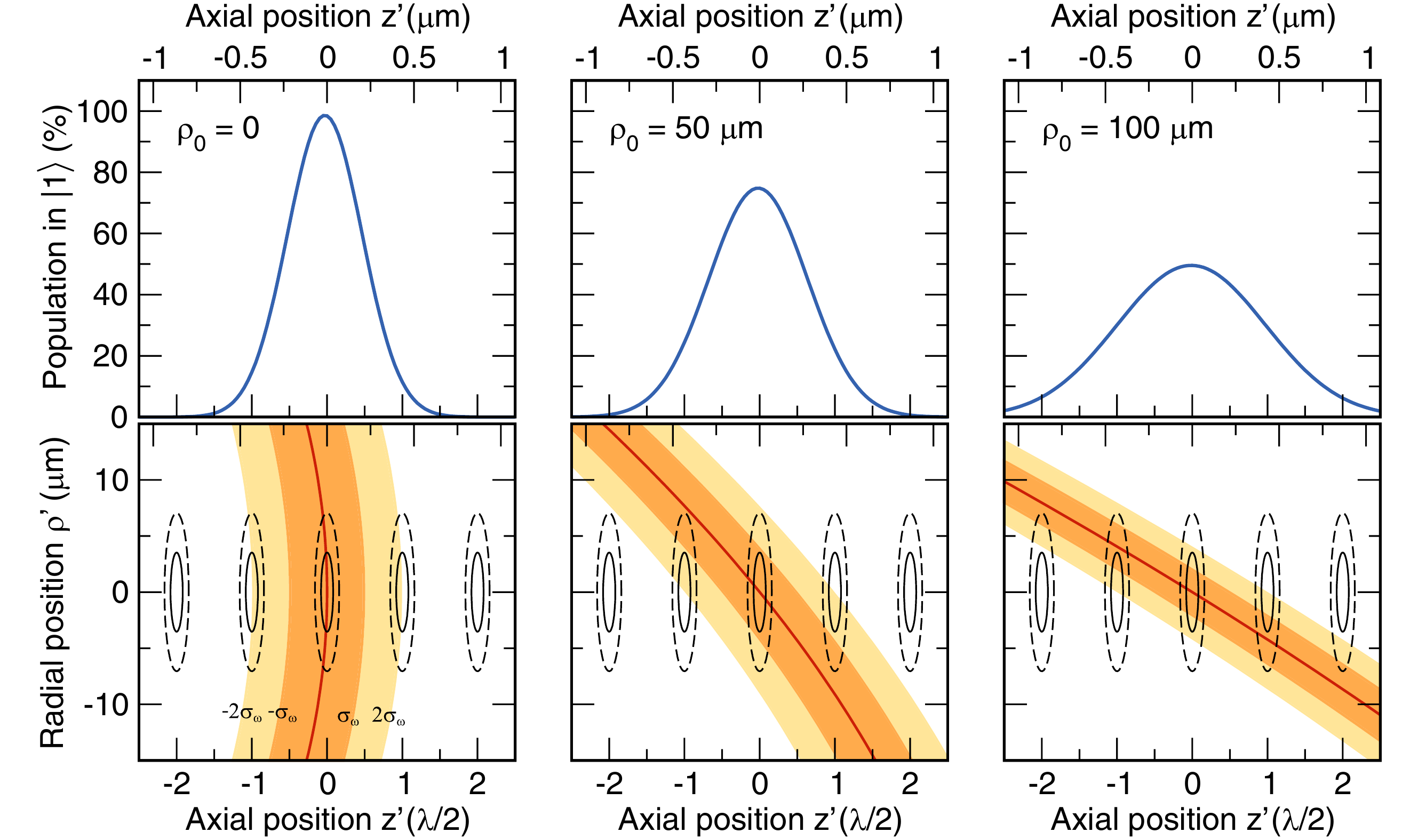}
	\caption{Addressed regions, defined by ${|\delta(\mathbf{r}^\prime, \mathbf{r}_{0})|\leq\sigma_{\omega}}$ and ${|\delta(\mathbf{r}^\prime, \mathbf{r}_{0})|\leq 2\sigma_{\omega}}$ in presence of the field gradient for different values of radial offsets $\rho_{0}$ (lower contour plots). The axes are expressed in units of lattice periodicity $\lambda/2$ and half of the beam waist $w_{0}/2$. Ellipses indicate the $1/\sqrt{e}$ (solid lines) and $1/e^{2}$ (dashed lines) spread of the Gaussian thermal wave packets of atoms trapped in the potential wells (sites) of the optical lattice. Upper graphs show the corresponding effective spectra in position space, i.e.~the population transfer as a function of axial displacement $z$ from the resonance position ${\mathbf{r}=0}$ for a Gaussian $\pi$-pulse with $\sigma_{\omega}/2\pi=6.4\,\mathrm{kHz}$ and ${P_{\mathrm{max}}=100\%}$.}\label{fig:selectivity-offset}

\end{figure*}
In Figure \ref{fig:selectivity-offset}, we show the calculated regions addressed by the pulse (addressed regions) for which $|\delta(\mathbf{r}^\prime, \mathbf{r}_{0})|\leq\sigma_{\omega}$ and $|\delta(\mathbf{r}^\prime, \mathbf{r}_{0})|\leq 2\sigma_{\omega}$ in presence of the field gradient for different values of radial offset $\rho_0$. Atoms located in these regions experience a minimum population transfer of ${P_{\mathrm{min}}\geq 0.6 P_{\mathrm{max}}}$ and ${P_{\mathrm{min}}\geq 0.13 P_{\mathrm{max}}}$, respectively, whenever the Gaussian $\pi$-pulse is applied. Because of the quadratical dependence of $\delta(\mathbf{r}^\prime, \mathbf{r}_{0})$ on the radial component, see Eqs.~(\ref{eq:detuning-field-gradient}) and (\ref{eq:transition-frequency-gradient}), the addressed regions are extending over neighboring lattice sites for increasing values of $\rho_0$. This yields a decrease of the spacial selectivity of the Gaussian $\pi$-pulses. Furthermore, we observe a decreasing intersection of the Gaussian thermal wave packet of a trapped atom and the addressed region. This leads to a decreased effective population transfer at desired lattice sites. We quantify both observations by calculating the expected position dependent spectra, i.\,e.\ the population transfer as function of axial displacement $z^\prime$ from the resonance position ${\mathbf{r}^\prime=0}$ for different radial offsets. For this, we average the population transfer specified by the spectrum of the Gaussian $\pi$-pulse (Eq.~(\ref{eq:gauss-pulse-spectrum})) over all position-dependent detunings, taking into account the axial and radial width of the atomic thermal wave packet  in the trapping potential of the lattice.

%
%

This yields again an approximately Gaussian shaped spectrum in position space, see Fig.~\ref{fig:selectivity-offset}, from which the \emph{effective} maximum population transfer $\bar{P}_{\mathrm{max}}$ and the \emph{effective} $1/\sqrt{e}$ half-width in position space $\bar{\sigma}_{z}$ can be inferred.

%

By comparing our measured spectrum to the calculated effective spectra, we deduce a calculated offset of $\rho_0 = (64 \pm 14)\,\mu$m. For this offset, we expect a maximum population transfer of ${\bar{P}_{\mathrm{max}}=(56\pm 8)\%}$, and thus, an expected number of correctly prepared atom pairs of $\bar{N}_{\mathrm{exp}} \approx 40\,$. This value agrees with the measured number of correctly prepared pairs stated above.

Moreover, by changing the homogeneous offset field $\mathbf{B}_0$, it is possible to infer and to reduce the offset between magnetic gradient and optical lattice axes, see Fig.~\ref{fig:RadialAxialShift}c. For this we record the spatial position of the microwave resonance along the gradient as a function of the radial offset $\rho_0$. This offset is increased by applying a homogeneous field along the $x^\prime$-direction. We observe the expected quadratic dependence of the resonance position. Adjusting the magnetic field offset to the measured value of the parabola maximum, the offset along the scanning direction can be removed. Repeating this for the orthogonal direction we are able to remove this offset completely from our system.

This effect is also important for the selection of 2D planes from a perfect 3D Mott-insulting state with unity filling. We estimate that for a spherical symmetric cloud of $25\,\mu$m diameter, for ideal alignment and with our lattice, gradient and pulse parameters, a 2D plane of 2500 atoms could be prepared by applying the inner loop of Fig.~\ref{fig:multi-push-out} twice. Here, approximately $3\%$ of the atoms stem from neighboring planes.

\section{Conclusion and Outlook}
We have created arbitrary patterns of atoms trapped in a 1D optical lattice with single-site resolution using magnetic resonance techniques. Such patterns can be used as initial configurations for various purposes, including collisional interaction between nearest neighbors. The mechanisms limiting the resolution have been identified as being technical in nature and can be overcome. Specifically, the optical lattice potential can be actively stabilized to avoid long term drifts, the magnetic field gradient can be increased and the microwave pulses can be shaped by methods of optimal control to increase the selectivity of the pulses while preserving maximal transfer probability. In a further extension of this method, arbitrary single qubit operations with single-site resolution - even in massively parallel operation - may become possible by taking into account off-resonant phase shifts on other qubits.

\section*{Acknowledgements}
We acknowledge financial support by a DFG research unit (FOR 635) and the IP SCALA. M.~K.~acknowledges partial support by the Studienstiftung des deutschen Volkes, J.-M.~C. received partial support from the Korea Research Foundation grant funded by the Korean Government (Ministry of Education and Human Resources Development).

\end{document}